# Write asymmetry of spin-orbit torque memory induced by in-plane magnetic fields

Baiqing Jiang, Dongyang Wu, Qianwen Zhao, Kaihua Lou, Yuelei Zhao, Yan Zhou, C. Tian, and Chong Bi

***Abstract*—Write asymmetry, the significantly different write current for high-to-low and low-to-high resistance switching because of natural stochastic behaviors of magnetization, is a fundamental issue in magnetic random-access memory (MRAM). For high-performance spin transfer torque (STT) MRAM, it can be eliminated by precisely controlling atomically thin magnetic multilayers or by introducing compensation techniques in circuit-level designs, while for spin-orbit torque (SOT) MRAM, it has not been addressed. Here we systematically investigated the write asymmetry of SOT-MRAM as a function of applied magnetic fields (H) and demonstrated that the write currents are intrinsically asymmetric due to different SOT efficiencies for high-to-low and low-to-high switching. Furthermore, we found that the SOT efficiency is very sensitive to the tilt angle between H and write current, which can be tuned through H to achieve symmetric SOT switching. These results provide an additional guideline for designing SOT devices and suggest that the write asymmetry can be eliminated by adjusting the introduced effective magnetic fields within a field-free SOT-MRAM architecture.**

***Index Terms*—MRAM, SOT, STT, nonvolatile memory**

## I. Introduction

MRAM based on perpendicularly magnetized FM/IL/FM (FM: ferromagnetic layer; IL: ultrathin insulating layer) tunnel junctions (MTJs) has attracted intense attentions from both academia and industry [1]–[3] owing to its potential applications in the replacement of static random-access memory (SRAM) when SRAM approaches its scaling limit at the 7-nm technology node and beyond [4], [5]. In addition to the excellent scalability compared to SRAM, MRAM shows low-power consumption because of its nonvolatility that does not need static power to retain information. Based on write mechanisms [6], MRAM can be classified into STT-MRAM and SOT-MRAM. The former uses the STT generated from the reference layer of MTJs to switch the storage layer [1], [3], while the latter uses the SOT generated by a neighboring SOT layer [7], [8]. In STT-MRAM, the critical switching current ($I_c$) as well as switching time varies when the resistance switches from high-to-low and low-to-high states, causing the significant write asymmetry [9]–[11]. This write asymmetry induces extra complexity during STT-MRAM production both in device- and circuit-level [11]–[13]. Combined with circuit compensation techniques, the write asymmetry of STT-MRAM can be eased by precisely controlling MTJ structures [1], [2].

Since SOT-MRAM is still in its infancy, most studies are focusing on the fundamental SOT phenomena such as field-free switching [14]–[18] and enhancement of SOT efficiency [19]–[24]. The write asymmetry issue has not been specifically raised so far even many works show asymmetric $I_c$ like STT [8], [25]. In the SOT scheme, an external H collinear with write current must be applied to break the symmetry for deterministic switching [26]. However, in practical devices, the external H cannot be aligned with the write current exactly and a small angle between them is usually required for achieving full SOT switching [7]. This symmetry-breaking and field misalignment may lead to different switching processes or SOT efficiencies

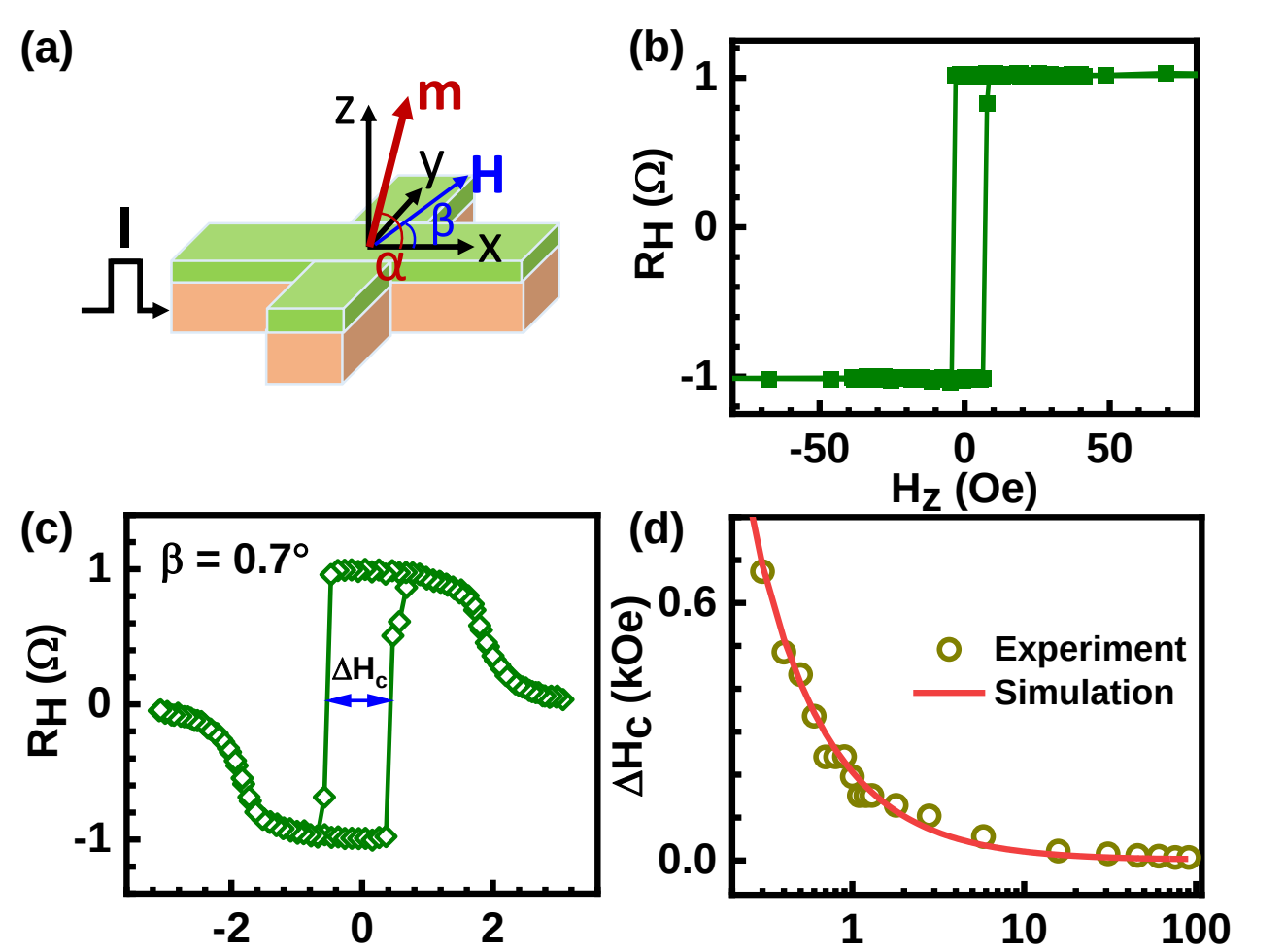

Fig. 1. (a) Illustration of SOT-devices and the measurement configurations. (b) The perpendicularly magnetized hysteresis loop. (c) $R_H$ as a function of in-plane field when β = 0.7°, where $\Delta H_c$ is defined. (d) The extracted $\Delta H_c$ as a function of β. The solid line is the simulated result based on the coherent switching.

Manuscript received July 17, 2021. This work is supported by the National Key R&D Program of China (Grant No. 2019YFB2005800 and 2018YFA0701500), the National Natural Science Foundation of China (Grant No. 61974160, 11974298, 61961136006, 61821091, and 61888102), and the Strategic Priority Research Program of the Chinese Academy of Sciences (Grant No. XDB44000000). Y. Z. acknowledges the support by Shenzhen Fundamental Research Fund (Grant No. JCYJ20210324120213037), Guangdong Special Support Project (2019BT02X030), Shenzhen Peacock Group Plan (KQTD20180413181702403), and Pearl River Recruitment Program of Talents (2017GC010293). *(Corresponding authors: Yan Zhou and Chong Bi.)*

Baiqing Jiang, Dongyang Wu, Qianwen Zhao, Kaihua Lou, and Chong Bi are with Institute of Microelectronics, Chinese Academy of Sciences, Beijing 100029, China

Yuelei Zhao is with The Chinese University of Hong Kong, Shenzhen, Guangdong 518172, China, and University of Science and Technology of China, Hefei, Anhui 230026, China.

Yan Zhou is with The Chinese University of Hong Kong, Shenzhen, Guangdong 518172, China (e-mail: zhouyan@cuhk.edu.cn).

C. Tian is with QSpinTech, Ltd., Beijing, 100020, China.

Chong Bi is with University of Chinese Academy of Sciences, Beijing 100049, China (e-mail: bichong@ime.ac.cn).

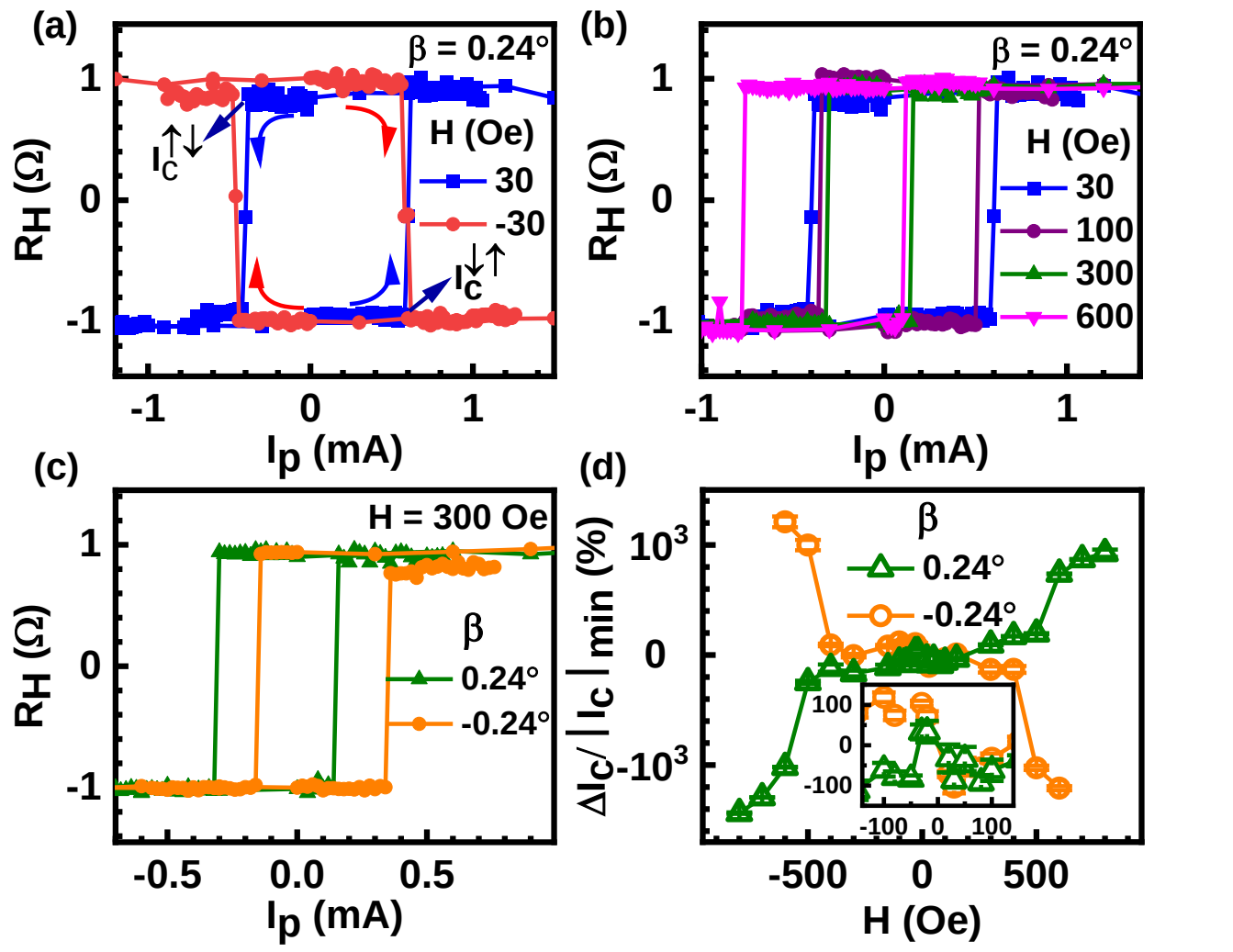

Fig. 2. (a) Typical current-driven SOT switching loops under ±H when β = 0.24º. (b) SOT switching loops under different H when β = 0.24º. (c) SOT switching loops under ±β when H = 300 Oe. (d) $\Delta I_c/|I_c|_{min}$ as a function of H. The inset is the enlarged part of the center region.

for up-to-down (↑↓) and down-to-up (↓↑) switching (corresponding to the low-to-high and high-to-low resistance switching of MTJs, respectively).

Moreover, two different SOT switching mechanisms, coherent switching [26], [27] and domain expansion switching [28]–[30], may dominate the ↑↓ and ↓↑ switching processes, respectively, also resulting in the write asymmetry. It should be noted that, in the field-free SOT switching architectures, an effective in-plane magnetic field due to stray fields or interlayer coupling can be viewed as the external H and the write asymmetry should also exist. In this work, we directly address the write asymmetry issue in typical SOT structures and explore the underlying physical origins. Finally, we show that the write asymmetry of SOT-MRAM can be completely eliminated by adjusting the external H.

## II. DEVICE FABRICATION

The samples for investigating SOT switching are reduced MTJ structures in the absence of reference layers, specifically, Ta 8/CoFeB 1.2/MgO 2 (nm) multilayers with a 3 nm $TaO_x$ capping layer. All layers were deposited on a thermally oxidized $Si/SiO_2$ 300 nm substrate by magnetron sputtering with a base pressure about 4.5 × $10^{-9}$ Torr. The Ta and CoFeB layers were sputtered from the Ta and $Co_2Fe_6B_2$ targets (purity: 99.95%) through DC sputtering under an Ar pressure of 2 mTorr, respectively. The MgO layer was sputtered from an MgO target (99.99%) through RF sputtering under the Ar pressure of 1.1 mTorr. The as-deposited layers were then transferred to a 220 °C vacuum annealing chamber with the pressure better than 1 × $10^{-7}$ Torr for one hour to enhance perpendicular magnetic anisotropy (PMA). The samples were patterned into Hall bars with the width of 2 μm through standard photolithography and subsequent Ar ion milling. As illustrated in Fig. 1(a), the up and down magnetization states were detected by measuring anomalous Hall resistance ($R_H$). The sensing current for measuring $R_H$ is 25 μA. For current-driven SOT switching, a 1 ms current pulse ($I_p$) was applied first to switch

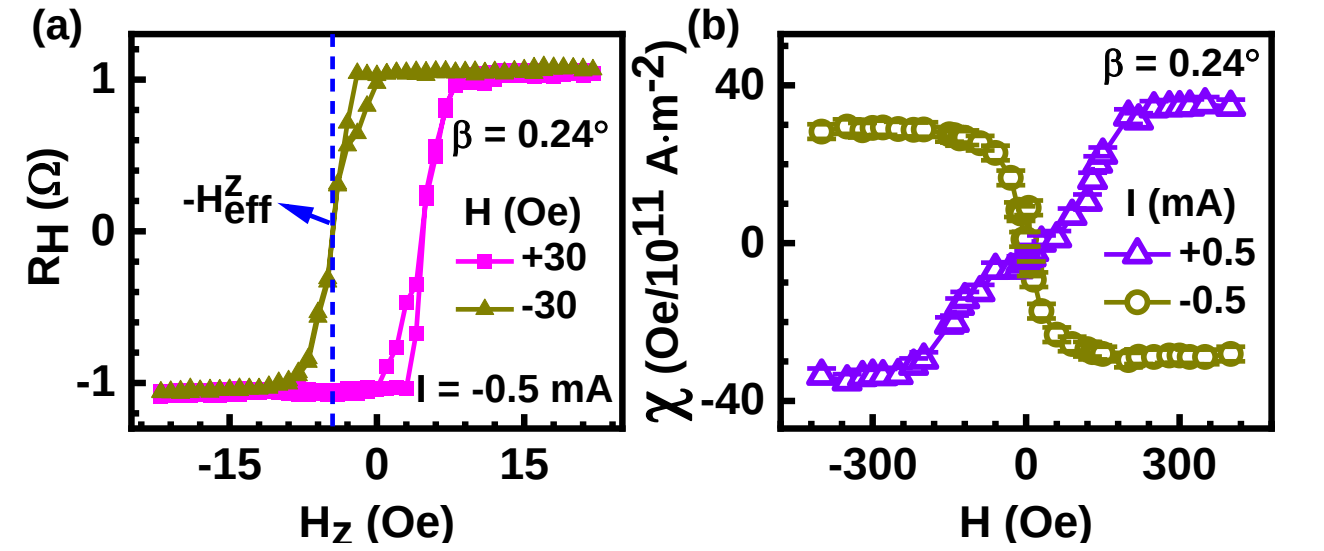

Fig. 3. (a) Perpendicularly magnetized $R_H$ loops under an in-plane field H = ±30 Oe with β = 0.24º. The shift of $R_H$ loops corresponds to $H_{eff}^z$. (b) The measured χ as a function of H when β = 0.24º.

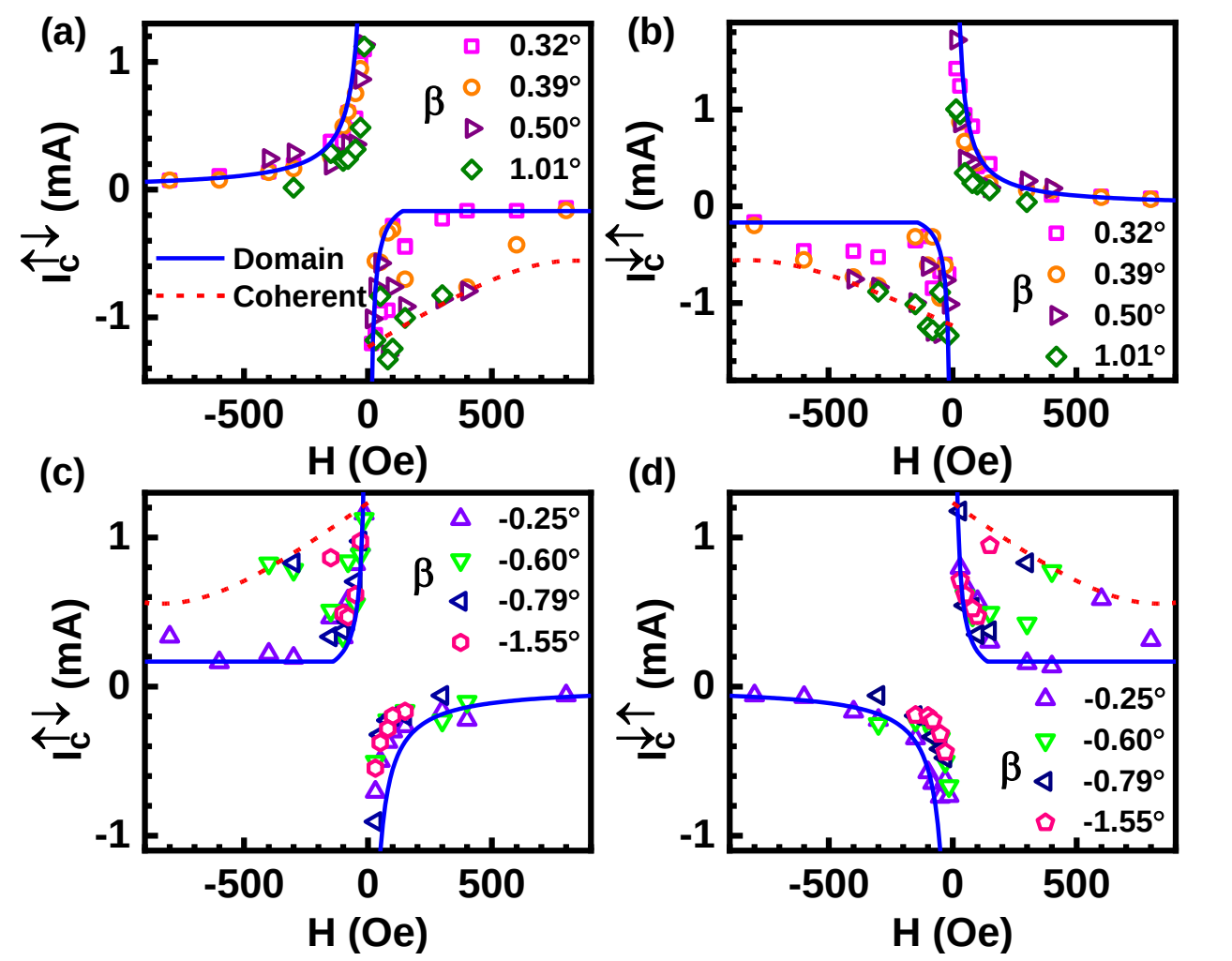

Fig. 4. The extracted $I_c$ for (a, c) ↑↓ and (b, d) ↓↑ switching with (a, b) positive and (c, d) negative β. The solid and dash lines represent the predicted values for the domain expansion switching and coherent switching, respectively. The parameters are $\varepsilon H_k$ = 0.15 T, $\beta$ = 0.3º, $\delta M_s$ = 5 × $10^4$ A/m, and $\theta_{SH}$ = 0.2.

the CoFeB layer and then the sensing current was applied to detect the magnetization states [23].

## III. SOT SWITCHING RESULTS AND DISCUSSION

The PMA was confirmed by measuring $R_H$ versus $H_z$, where a square switching loop is clearly presented (Fig. 1(b)). The critical switching field ($H_s^z$) is about ± 5 Oe, which can be used to estimate the tilt angle (β) of H from film plane through $\Delta H_c \sin\beta = 2H_s^z$, where $\Delta H_c$ is defined in Fig. 1(c). Fig. 1(c) shows typical $R_H$ versus H curve when β = 0.7º, from which $H_k$ ≈ 2.2 kOe is extracted according to the field-driven coherent switching model [23], [26]. Fig. 1(d) presents $\Delta H_c$ as a function of β extracted from Fig. 1(c), where the excellent agreement of the simulated coherent switching behaviors with experimental results indicates that the field-driven coherent switching keeps well when β ≥ 0.2º.

Fig. 2(a) and Fig. 2(b) show typical current-induced SOT switching curves when β = 0.24°, in which $I_c$ for ↑↓ ($I_c^{\uparrow\downarrow}$) and ↓↑ ($I_c^{\downarrow\uparrow}$) switching can be extracted. For +H and -H, the switching loops are anticlockwise and clockwise, respectively, which are the typical characteristics of SOT switching [7], [8], [26]. The remarkable feature is that the ↑↓ and ↓↑ switching becomes asymmetric with increasing H (Fig. 2(b)). When H = 300 Oe, $I_c^{\uparrow\downarrow}$ = -0.31 mA, two times larger than $I_c^{\downarrow\uparrow}$= 0.14 mA. The $I_c$ difference becomes even larger when H = 600 Oe. Fig. 2(c) gives the SOT switching loops when β = ± 0.24°, which clearly shows that the asymmetric SOT switching strongly depends on the sign of β. For β = 0.24°, $|I_c^{\uparrow\downarrow}| > |I_c^{\downarrow\uparrow}|$, while for β = -0.24°, $|I_c^{\uparrow\downarrow}| < |I_c^{\downarrow\uparrow}|$.

To clearly demonstrate the asymmetric SOT switching, $\Delta I_c/|I_c|_{min}$ with $\Delta I_c = |I_c^{\uparrow\downarrow}| - |I_c^{\downarrow\uparrow}|$ as a function of H was plotted in Fig. 2(d), where $|I_c|_{min} = \text{Min}\{|I_c^{\uparrow\downarrow}|, |I_c^{\downarrow\uparrow}|\}$. When $|H|$> 400 Oe, $\Delta I_c$ is determined by the sign of β and $|\Delta I_c|/|I_c|_{min}$ exceeds 1000%, showing strongly asymmetric switching. Even for weak fields $|H|$< 100 Oe, $|\Delta I_c|/|I_c|_{min}$ can be larger than 100%. Remarkably, $\Delta I_c$ can approach 0 at certain fields (for example, H ≈ ±200 Oe), which means that the asymmetric SOT switching can be eliminated by controlling H. It should be noted that, when $|\beta|$ < 0.24º, both field- and current-driven magnetization switching only show partial switching and $I_c$ cannot be determined accurately.

We have also evaluated the SOT efficiency by measuring the SOT-induced effective perpendicular field ($H_{eff}^z$) [31], [32] to explore the physical origins of the asymmetric SOT switching. $H_{eff}^z$ originates from SOT effects on the internal magnetization texture of Néel-type domain walls and directly represents SOT efficiency [31]. Fig. 3(a) shows $R_H$ versus $H_z$ loops with an additional in-plane magnetic field H (β = 0.24º), where the shift of $R_H$ loops can be used to estimate $H_{eff}^z$. For the additional H = 30 Oe, the $R_H$ loop shifts toward $+H_z$, indicating a negative $H_{eff}^z$ generated by the -0.5 mA applied current. The negative $H_{eff}^z$ will induce the magnetization switching to the down state when $|H_{eff}^z| > |H_s^z|$, in consistent with the SOT switching directions in Fig. 2(a).

Fig. 3(b) gives the measured SOT efficiency ($\chi = H_{eff}^z/|J|$) as a function of H, where *J* is the applied current density. Both the saturation field (corresponding to the effective Dzyaloshinskii-Moriya interaction (DMI) field ($H_{DMI}$)) and $\chi$ agree well with previous reports [32]. Interestingly, χ shows different field dependence for opposite currents in the linear range when H approaches 0 Oe. For +0.5 mA, the slope of χ in the linear range ($|H| < 150$ Oe) is about 0.14 × $10^{-11}$ $m^2$/A and the estimated $H_{DMI}$ ≈ 240 Oe, while for -0.5 mA, the slope in the linear range ($|H| < 60$ Oe) is about -0.54 × $10^{-11}$ $m^2$/A and $H_{DMI}$ ≈ 100 Oe. Moreover, the saturation value of χ also shows a slight difference for opposite currents. We have also measured χ when β = -0.24º and observed a reversal χ behavior for ±0.5 mA, indicating that the significantly different χ dependence is due to β.

This remarkable current direction dependence of SOT efficiency has not been reported before probably because χ was assumed to be the same for ±I [32]. Moreover, the strong β-dependence of χ may explain the significant χ change in the wedged samples (< 0.001 nm) [32] where β changes slightly compared to the uniformly deposited samples even the thickness variation may be less than one atomic layer in a single wedged device [17], [32]. According to the 1D domain wall model, $\chi = (\hbar\pi\xi_{DL}/4e\mu_0 M_s t_{FM})\cos\Phi$, where $\xi_{DL}$, $M_s$, $t_{FM}$, and Φ, are the damping-like SOT, the saturation magnetization and the thickness of the ferromagnetic layer, and the tilt angle of domain wall moment, respectively. When H << $H_{DMI}$, cosΦ ≈ $H/H_{DMI}$, it is reasonable that a larger χ slope corresponds to a smaller $H_{DMI}$, as observed in Fig. 3(b). However, the saturation value, $\chi_{sat} = \hbar\pi\xi_{DL}/4e\mu_0 M_s t_{FM}$, varying with current direction cannot be fully understood within the 1D domain wall model [31], indicating that the detailed dynamics of the internal magnetization texture during SOT-driven domain wall motion should be considered.

Fig. 4 shows the measured $I_c$ as a function of H. The solid lines guide $I_c$ for the magnetization switching induced by the domain expansion, which are determined from the measured χ by using $H_{eff}^z = H_s^z$ [33]. Overall, $I_c$ increases sharply with reducing H and gradually decreases when $|H|$ > 100 Oe, in contrast to the coherent switching where $I_c$ shows a linear dependence on H [27]. For +β, a positive $I_c$ (for example, in the upper left corner of Fig. 4(a)) is always smaller than the value of a negative $I_c$ (in the bottom right corner of Fig. 4(a)) in the field range of $|H|$ > 240 Oe, in consistent with the χ measurements (Fig. 3(b)), where $H_{eff}^z$ for +I is larger than that for -I when $|H|$ > $H_{DMI}$ (≈ 240 Oe). These results confirms that the SOT switching strongly correlates with the measured χ and thus is dominated by the domain expansion. By considering χ is very sensitive to β, the symmetric switching can be achieved by precisely controlling H and film roughness to make $|\chi(+I)| = |\chi(-I)|$.

Notably, $I_c$ occasionally shows extremely large values which has a linear dependence on H, as predicted by the coherent switching [27]. With the involvement of β, $I_c$ of the coherent switching can be developed as,

$$I_c = \left(\frac{e\delta\varepsilon M_s H_k t_{FM} S}{\hbar\theta_{SH}}\right)\sqrt{1 + 2\sqrt{2}(\sin\beta - \cos\beta)h_x + (2.5 - 3\sin\beta\cos\beta)h_x^2},$$

where $\theta_{SH}$ is the effective spin Hall angle, $S$ is the cross-sectional area of the SOT layer, $h_x = H/\varepsilon H_k$, $0 \le \delta, \varepsilon \le 1$ are the factors for describing the residual $M_s$ and $H_k$ due to Joule heating, respectively. The dash lines in Fig. 4 represent $I_c$ for the coherent switching. The overlap with the occasional large values indicates that the coherent switching can also happen randomly, which will induce the additional write asymmetry.

## IV. Conclusion

We have addressed the asymmetric switching issue in the prototype SOT-MRAM devices by varying the applied external field. It is found that the SOT efficiencies are intrinsically different for the up-to-down and down-to-up switching, which makes the corresponding SOT switching asymmetric in principle. The combination of the measured SOT efficiency and critical SOT switching current further reveals that the magnetization switching is dominated by the domain expansion. These results provide basic guidelines for designing symmetric SOT switching devices. Furthermore, we expect that the asymmetric SOT switching mechanism still functions in the nano-sized SOT devices since the width of domain walls is only several nanometers and the magnetization switching is still dominated by the domain expansion.

## REFERENCES

[1] S. Bhatti, R. Sbiaa, A. Hirohata, H. Ohno, S. Fukami, and S. N. Piramanayagam, “Spintronics based random access memory: a review,” *Mater. Today*, vol. 20, no. 9, pp. 530–548, Nov. 2017, doi: 10.1016/J.MATTOD.2017.07.007.

[2] S. Ikegawa, F. B. Mancoff, J. Janesky, and S. Aggarwal, “Magnetoresistive Random Access Memory: Present and Future,” *IEEE Trans. Electron Devices*, vol. 67, no. 4, pp. 1407–1419, Apr. 2020, doi: 10.1109/TED.2020.2965403.

[3] Z. Zhang, Z. Wang, T. Shi, C. Bi, F. Rao, Y. Cai, Q. Liu, H. Wu, and P. Zhou, “Memory materials and devices: From concept to application,” *InfoMat*, vol. 2, no. 2, pp. 261–290, Mar. 2020, doi: 10.1002/INF2.12077.

[4] F. Oboril, R. Bishnoi, M. Ebrahimi, and M. B. Tahoori, “Evaluation of hybrid memory technologies using SOT-MRAM for on-chip cache hierarchy,” *IEEE Trans. Comput. Des. Integr. Circuits Syst.*, vol. 34, no. 3, pp. 367–380, Mar. 2015, doi: 10.1109/TCAD.2015.2391254.

[5] S. Sakhare, M. Perumkunnil, T. H. Bao, S. Rao, W. Kim, D. Crotti, F. Yasin, S. Couet, J. Swerts, S. Kundu, D. Yakimets, R. Baert, H. Oh, A. Spessot, A. Mocuta, G. S. Kar, and A. Furnemont, “Enablement of STT-MRAM as last level cache for the high performance computing domain at the 5nm node,” in *IEDM Tech. Dig.*, 2018, pp. 18.3.1-18.3.4. doi: 10.1109/IEDM.2018.8614637.

[6] A. Brataas, A. D. Kent, and H. Ohno, “Current-induced torques in magnetic materials,” *Nat. Mater.*, vol. 11, no. 5, pp. 372–381, 2012, doi: 10.1038/NMAT3311.

[7] I. M. Miron, K. Garello, G. Gaudin, P. J. Zermatten, M. V. Costache, S. Auffret, S. Bandiera, B. Rodmacq, A. Schuhl, and P. Gambardella, “Perpendicular switching of a single ferromagnetic layer induced by in-plane current injection,” *Nature*, vol. 476, no. 7359, pp. 189–193, Aug. 2011, doi: 10.1038/NATURE10309.

[8] L. Liu, C. F. Pai, Y. Li, H. W. Tseng, D. C. Ralph, and R. A. Buhrman, “Spin-torque switching with the giant spin hall effect of tantalum,” *Science (80-. ).*, vol. 336, no. 6081, pp. 555–558, May 2012, doi: 10.1126/SCIENCE.1218197.

[9] R. Bishnoi, M. Ebrahimi, F. Oboril, and M. B. Tahoori, “Improving Write Performance for STT-MRAM,” *IEEE Trans. Magn.*, vol. 52, no. 8, p. 3401611, Aug. 2016, doi: 10.1109/TMAG.2016.2541629.

[10] T. Devolder, J. Hayakawa, K. Ito, H. Takahashi, S. Ikeda, P. Crozat, N. Zerounian, J.-V. Kim, C. Chappert, and H. Ohno, “Single-Shot Time-Resolved Measurements of Nanosecond-Scale Spin-Transfer Induced Switching: Stochastic Versus Deterministic Aspects,” *Phys. Rev. Lett.*, vol. 100, no. 5, p. 057206, Feb. 2008, doi: 10.1103/PhysRevLett.100.057206.

[11] K. Ikegami, H. Noguchi, C. Kamata, M. Amano, K. Abe, K. Kushida, E. Kitagawa, T. Ochiai, N. Shimomura, S. Itai, D. Saida, C. Tanaka, A. Kawasumi, H. Hara, J. Ito, and S. Fujita, “Low power and high density STT-MRAM for embedded cache memory using advanced perpendicular MTJ integrations and asymmetric compensation techniques,” in *IEDM Tech. Dig.*, 2014, pp. 28.1.1-28.1.4. doi: 10.1109/IEDM.2014.7047123.

[12] W. S. Zhao, T. Devolder, Y. Lakys, J. O. Klein, C. Chappert, and P. Mazoyer, “Design considerations and strategies for high-reliable STT-MRAM,” *Microelectron. Reliab.*, vol. 51, no. 9–11, pp. 1454–1458, Sep. 2011, doi: 10.1016/J.MICROREL.2011.07.001.

[13] S. H. Choday, S. K. Gupta, and K. Roy, “Write-optimized STT-MRAM bit-cells using asymmetrically doped transistors,” *IEEE Electron Device Lett.*, vol. 35, no. 11, pp. 1100–1102, Nov. 2014, doi: 10.1109/LED.2014.2358998.

[14] Y. W. Oh, S. H. C. Baek, Y. M. Kim, H. Y. Lee, K. D. Lee, C. G. Yang, E. S. Park, K. S. Lee, K. W. Kim, G. Go, J. R. Jeong, B. C. Min, H. W. Lee, K. J. Lee, and B. G. Park, “Field-free switching of perpendicular magnetization through spin-orbit torque in antiferromagnet/ferromagnet/oxide structures,” *Nat. Nanotechnol.*, vol. 11, no. 10, pp. 878–884, Oct. 2016, doi: 10.1038/NNANO.2016.109.

[15] S. Fukami, C. Zhang, S. Duttagupta, A. Kurenkov, and H. Ohno, “Magnetization switching by spin-orbit torque in an antiferromagnet-ferromagnet bilayer system,” *Nat. Mater.*, vol. 15, no. 5, pp. 535–541, May 2016, doi: 10.1038/NMAT4566.

[16] T. Simsek, “Field-Free Spin-Orbit Torque Switching in Magnetic Tunnel Junction Structures with Stray Fields,” *IEEE Magn. Lett.*, vol. 12, p. 4500305, 2021, doi: 10.1109/LMAG.2021.3063081.

[17] G. Yu, P. Upadhyaya, Y. Fan, J. G. Alzate, W. Jiang, K. L. Wong, S. Takei, S. A. Bender, L.-T. Chang, Y. Jiang, M. Lang, J. Tang, Y. Wang, Y. Tserkovnyak, P. K. Amiri, and K. L. Wang, “Switching of perpendicular magnetization by spin–orbit torques in the absence of external magnetic fields,” *Nat. Nanotechnol. 2014 97*, vol. 9, no. 7, pp. 548–554, May 2014, doi: 10.1038/nnano.2014.94.

[18] K. Garello, F. Yasin, H. Hody, S. Couet, L. Souriau, S. H. Sharifi, J. Swerts, R. Carpenter, S. Rao, W. Kim, J. Wu, K. K. V. Sethu, M. Pak, N. Jossart, D. Crotti, A. Furnemont, and G. S. Kar, “Manufacturable 300mm platform solution for Field-Free Switching SOT-MRAM,” in *Proc. Symp. VLSI Circuits*, Jun. 2019, pp. T194–T195. doi: 10.23919/VLSIC.2019.8778100.

[19] M. DC, R. Grassi, J.-Y. Chen, M. Jamali, D. R. Hickey, D. Zhang, Z. Zhao, H. Li, P. Quarterman, Y. Lv, M. Li, A. Manchon, K. A. Mkhoyan, T. Low, and J.-P. Wang, “Room-temperature high spin–orbit torque due to quantum confinement in sputtered Bi x Se (1–x) films,” *Nat. Mater.*, vol. 17, no. 9, pp. 800–807, Jul. 2018, doi: 10.1038/s41563-018-0136-z.

[20] W. Wang, T. Wang, V. P. Amin, Y. Wang, A. Radhakrishnan, A. Davidson, S. R. Allen, T. J. Silva, H. Ohldag, D. Balzar, B. L. Zink, P. M. Haney, J. Q. Xiao, D. G. Cahill, V. O. Lorenz, and X. Fan, “Anomalous spin–orbit torques in magnetic single-layer films,” *Nat. Nanotechnol.*, vol. 14, no. 9, pp. 819–824, Sep. 2019, doi: 10.1038/S41565-019-0504-0.

[21] L. Zhu, D. C. Ralph, and R. A. Buhrman, “Enhancement of spin transparency by interfacial alloying,” *Phys. Rev. B*, vol. 99, no. 18, p. 180404(R), May 2019, doi: 10.1103/PhysRevB.99.180404.

[22] L. L. Y. L. H. W. T. D. C. R. R. A. B. Chi-Feng Pai, “Spin transfer torque devices utilizing the giant spin Hall effect of tungsten,” *Appl. Phys. Lett.*, vol. 101, no. 12, p. 122404, Sep. 2012, doi: 10.1063/1.4753947.

[23] C. Bi, C. Sun, M. Xu, T. Newhouse-Illige, P. M. Voyles, and W. Wang, “Electrical Control of Metallic Heavy-Metal-Ferromagnet Interfacial States,” *Phys. Rev. Appl.*, vol. 8, no. 3, p. 034003, Sep. 2017, doi: 10.1103/PhysRevApplied.8.034003.

[24] C.-F. Pai, Y. Ou, L. H. Vilela-Leão, D. C. Ralph, and R. A. Buhrman, “Dependence of the efficiency of spin Hall torque on the transparency of Pt/ferromagnetic layer interfaces,” *Phys. Rev. B*, vol. 92, no. 6, p. 064426, Aug. 2015, doi: 10.1103/PhysRevB.92.064426.

[25] T. C. Chuang, C. F. Pai, and S. Y. Huang, “Cr-induced Perpendicular Magnetic Anisotropy and Field-Free Spin-Orbit-Torque Switching,” *Phys. Rev. Appl.*, vol. 11, no. 6, p. 061005, Jun. 2019, doi: https://doi.org/10.1103/PhysRevApplied.11.061005.

[26] L. Liu, O. J. Lee, T. J. Gudmundsen, D. C. Ralph, and R. A. Buhrman, “Current-Induced Switching of Perpendicularly Magnetized Magnetic Layers Using Spin Torque from the Spin Hall Effect,” *Phys. Rev. Lett.*, vol. 109, no. 9, p. 096602, Aug. 2012, doi: 10.1103/PhysRevLett.109.096602.

[27] K. S. Lee, S. W. Lee, B. C. Min, and K. J. Lee, “Threshold current for switching of a perpendicular magnetic layer induced by spin Hall effect,” *Appl. Phys. Lett.*, vol. 102, no. 11, Mar. 2013, doi: 10.1063/1.4798288.

[28] O. J. Lee, L. Q. Liu, C. F. Pai, Y. Li, H. W. Tseng, P. G. Gowtham, J. P. Park, D. C. Ralph, and R. A. Buhrman, “Central role of domain wall depinning for perpendicular magnetization switching driven by spin torque from the spin Hall effect,” *Phys. Rev. B*, vol. 89, no. 2, p. 024418, Jan. 2014, doi: 10.1103/PhysRevB.89.024418.

[29] C. Bi, H. Almasi, K. Price, T. Newhouse-Illige, M. Xu, S. R. Allen, X. Fan, and W. Wang, “Anomalous spin-orbit torque switching in synthetic antiferromagnets,” *Phys. Rev. B*, vol. 95, no. 10, p. 104434, Mar. 2017, doi: 10.1103/PhysRevB.95.104434.

[30] G. Yu, P. Upadhyaya, K. L. Wong, W. Jiang, J. G. Alzate, J. Tang, P. K. Amiri, and K. L. Wang, “Magnetization switching through spin-Hall-effect-induced chiral domain wall propagation,” *Phys. Rev. B*, vol. 89, no. 10, p. 104421, Mar. 2014, doi: 10.1103/PhysRevB.89.104421.

[31] A. Thiaville, S. Rohart, É. Jué, V. Cros, and A. Fert, “Dynamics of Dzyaloshinskii domain walls in ultrathin magnetic films,” *EPL (Europhysics Lett.*, vol. 100, no. 5, p. 57002, Dec. 2012, doi: 10.1209/0295-5075/100/57002.

[32] C. F. Pai, M. Mann, A. J. Tan, and G. S. D. Beach, “Determination of spin torque efficiencies in heterostructures with perpendicular magnetic anisotropy,” *Phys. Rev. B*, vol. 93, no. 14, p. 144409, Apr. 2016, doi: 10.1103/PhysRevB.93.144409.

[33] T. Y. Chen, C. Te Wu, H. W. Yen, and C. F. Pai, “Tunable spin-orbit torque in Cu-Ta binary alloy heterostructures,” *Phys. Rev. B*, vol. 96, no. 10, p. 104434, 2017, doi: 10.1103/PhysRevB.96.104434.